\documentclass[aps,prd,twocolumn,preprintnumbers,superscriptaddress,dblfloatfix,nofootinbib]{revtex4-1}
\usepackage[utf8]{inputenc}
\usepackage{lmodern}
\usepackage{amsmath,amssymb,mhchem}
\usepackage{graphicx}
\usepackage{url}
\usepackage{color}
\usepackage{enumitem}
\usepackage{subfigure}
\usepackage[dvipsnames]{xcolor}
\usepackage[colorlinks=true,breaklinks=true]{hyperref}
\hypersetup{allcolors=[rgb]{0.0 0.0 0.6},linkcolor=[rgb]{0.75 0.05 0.05}}
\usepackage{bm}
\usepackage{epsfig}
\usepackage{amsmath}
\usepackage{amssymb}
\usepackage{slashed}
\usepackage{color}
\usepackage{accents}
\usepackage{url}
\usepackage[dvipsnames]{xcolor}

\newcommand{\exclude}[1]{}

\begin{document}

%\preprint{}

\title{Scalar Direct Detection: In-Medium Effects}

\author{Graciela B. Gelmini}\email{gelmini@physics.ucla.edu}
\affiliation{Department of Physics and Astronomy, University of California, Los Angeles \\ Los Angeles, California, 90095-1547, USA}

\author{Volodymyr Takhistov}\email{vtakhist@physics.ucla.edu}
\affiliation{Department of Physics and Astronomy, University of California, Los Angeles \\ Los Angeles, California, 90095-1547, USA}

\author{Edoardo Vitagliano}\email{edoardo@physics.ucla.edu}
\affiliation{Department of Physics and Astronomy, University of California, Los Angeles \\ Los Angeles, California, 90095-1547, USA}

\date{\today}

%%%%%%%%%%%%%%%%%%%%%%%%%%%%%%%%%%%
\begin{abstract}
A simple extension of the Standard Model consists of a scalar field that can potentially constitute the dark matter (DM). Significant attention has been devoted to probing light $\mathcal{O}(\lesssim 10~\rm{eV})$ scalar DM, with a multitude of experimental proposals based on condensed matter systems as well as novel materials.
However, the previously overlooked effective in-medium mixing of light scalars with longitudinal plasmons can strongly modify the original sensitivity calculations of the direct detection experiments. We implement the in-medium effects for scalar DM detection, using thermal field theory techniques, and show that the reach of a large class of direct DM detection experiments searching for light scalars is significantly reduced. This development identifies setups based on Dirac materials and tunable plasma haloscopes as particularly promising for scalar DM detection. Further, we also show that scalars with significant boost with respect to halo DM, such as those produced in the Sun, decay of other particles or by
cosmic rays, will not suffer from in-medium suppression. Hence, multi-tonne direct DM detection experiments, such as those based on xenon or argon, also constitute a favorable target. We also discuss scalar mediated DM scattering.
\end{abstract}
%%%%%%%%%%%%%%%%%%%%%%%%%%%%%%%%%%% 
\maketitle
%%%%%%%%%%%%%%%%%%%%%%%%%%%%%%%%%%%

\section{Introduction}

Unveiling the nature of dark matter (DM) remains one of the most important missions of particle physics, astrophysics and cosmology (see e.g. Ref.~\cite{Gelmini:2015zpa} for a review). The possible mass of the DM constituents spans an enormous range of about 19 orders of magnitude.  
While historically many DM searches have focused on electroweak 
mass scales, there has been a strong interest in recent years in exploring DM particle candidates significantly lighter than $\mathcal{O}$(GeV).

Additional scalar fields constitute among the simplest extensions of the Standard Model (SM). Light scalars can readily appear in a variety of well motivated models, e.g. as moduli~\cite{Burgess:2010sy,Cicoli:2011yy,Dimopoulos:1996kp}, dilatons~\cite{Damour:1994zq,Taylor:1988nw} or in
Higgs portal models~\cite{Piazza:2010ye}. The smallness of mass of a light scalar field can be attributed to an enhanced symmetry, such as a conformal symmetry or supersymmetry. In the early Universe, light scalars can be produced through the well known misalignment mechanism~(e.g.~\cite{Arias:2012az}).

Here, we consider light scalars, which could constitute the DM and/or could be produced from some source, e.g. the Sun or cosmic-ray interactions, 
and  could be detected through the energy they deposit within a detector. We remain agnostic about the particular mechanism
to make the scalar we consider light, as well as to the mechanism by which it could account for the whole of the DM. In most models scalars can typically couple to fermions, and we will proceed phenomenologically and consider that the scalar field is coupled to electrons (see e.g.~\cite{Adelberger_2003, Piazza:2010ye, Arvanitaki_2015, Arvanitaki_2016,  Hochberg:2016ajh, Hochberg:2017wce}). Later, we also consider light scalars as mediators of DM interactions with the SM.

Direct detection experiments attempt to measure the energy deposited within a detector by interactions of particles passing through it. These could be DM particles in the dark halo of our Galaxy, or particles emitted by particular sources. Halo DM  particles lighter than $\mathcal{O}$(GeV) could not deposit sufficient energy in interactions with nuclei in ton-scale direct detection experiments  (e.g.~\cite{Aprile_2019,Aprile:2019dbj,Agnes:2018oej, Akerib:2016vxi}), whose
energy thresholds are close to $\mathcal{O}~(\mathrm{keV})$. 
Hence, other types of searches are required to explore lighter DM particles.  

Sensitivity to  $\mathcal{O}$(10 eV) mass halo DM can be gained within large direct detection experiments searching for ionization signals due to DM absorption by bound electrons~\cite{An:2014twa,Bloch:2016sjj}. Detection proposals based on materials with a small  band-gap $\mathcal{O}$(meV) envision having good sensitivity to halo DM particles with mass larger than the gap, through the efficient excitation of quasi-particles, such as phonons, magnons, and plasmons. Proposed
target materials include superconductors~\cite{Hochberg:2015pha,Hochberg:2015fth}, graphene~\cite{Hochberg:2016ntt}, Dirac materials~\cite{Hochberg:2017wce,Coskuner:2019odd}, superfluid helium~\cite{Knapen:2016cue,Schutz:2016tid,Guo:2013dt,Acanfora:2019con,Caputo:2019cyg} and polar materials~\cite{Knapen:2017ekk}, as well as plasma haloscopes~\cite{Lawson:2019brd, Gelmini:2020kcu}. Proposals based on molecular targets have also been advanced~\cite{Arvanitaki:2017nhi}.

In-medium effects within detectors could suppress the interactions of incoming particle and thus present a major limiting factor for experimental searches. While this is well-known for searches of dark photons (as noticed in Ref.~\cite{An:2013yua}, see also e.g.~\cite{An:2014twa, Hochberg:2017wce} and, for a recent update,~\cite{An:2020bxd}), gauge vector bosons 
having a kinetic mixing with the SM photon~\cite{Holdom:1985ag}, this effect has been ignored in the literature when discussing scalar 
absorption or scalar-mediated DM detection. In this work, we show that the suppression of scalar interactions in a medium can significantly modify the detection sensitivity and thus be of paramount importance in  the design of  experiments.

 In the following, we will discuss how scalar absorption can be
 described with thermal field theory tools, and find that the mixing with the longitudinal plasmon affects the interaction of the scalar with the medium. We will then use superconductor detectors as an example in which in-medium effects are important. After briefly discussing scalar mediated DM scattering, we identify possible ways to avoid the in-medium suppression, which will bring us to our conclusions. The  Appendices  discuss the calculation of self-energies in a plasma using thermal field theory and the absorption of scalars in superconductors.  

\section{Scalar dark matter absorption}

Consider a light scalar field $\phi$ that will in general couple to the SM fermions, e.g.~via a Higgs-mixing portal~\cite{Piazza:2010ye}. The resulting low energy Lagrangian is
\begin{equation}\label{eq:lagrangian}
	\mathcal{L} \supset \frac{1}{2} (\partial_\mu \phi)^2 - \frac{1}{2} m_\phi^2 \phi^2
	+ \sum_f g_{\phi f} \phi \bar{f} f \, .
\end{equation}
Let us assume that $\phi$ couples to electrons through the interaction term $g\phi\bar{e}e$ and also that it constitutes the entirety of the DM.

In a plasma, $\phi$ can mix with the in-medium longitudinal plasmon. The latter is the quasi-particle of the longitudinal component of the electromagnetic field in a medium. Physically, this corresponds to a quantum of collective oscillation of the electron gas around the ions constituting the medium. 

We employ thermal field theory to obtain the scalar DM absorption rate in a detector. The main advantage of resorting to this approach instead of the kinetic theory approach is that mixing effects are explicit and straightforward to implement. 
In the kinetic theory approach, one computes, with usual quantum field theory techniques, the velocity averaged absorption rate $\langle n_e \sigma_{\rm abs}v_{\rm rel}\rangle$ 
where $n_e$ is the electron number number density and $\sigma_{\rm abs}$ is the absorption cross-section.  In-medium effects are
included in $\sigma_{\rm abs}$
as ``corrections" in an ad-hoc way following general arguments. Such procedure has been widely employed in the literature to treat in-medium effects in the context, for example, of~dark photon absorption~(e.g.~\cite{An:2013yua,An:2014twa, Hochberg:2017wce}).

 \begin{figure*}[tb]
\begin{center}
\includegraphics[trim={0mm 0mm 0 0},clip,width=.45\textwidth]{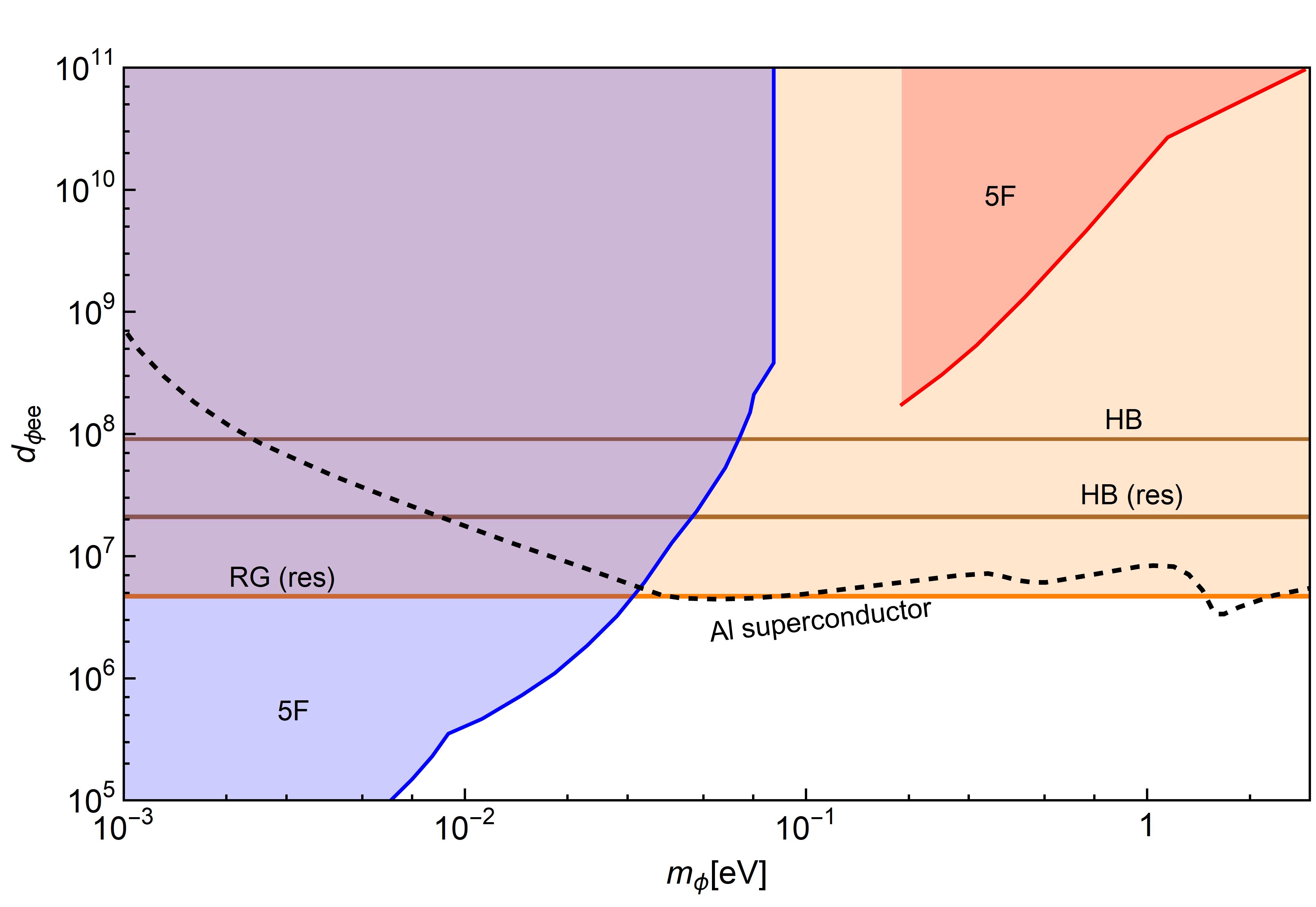}
\includegraphics[trim={0mm 0mm 0 0mm},clip,width=.45\textwidth]{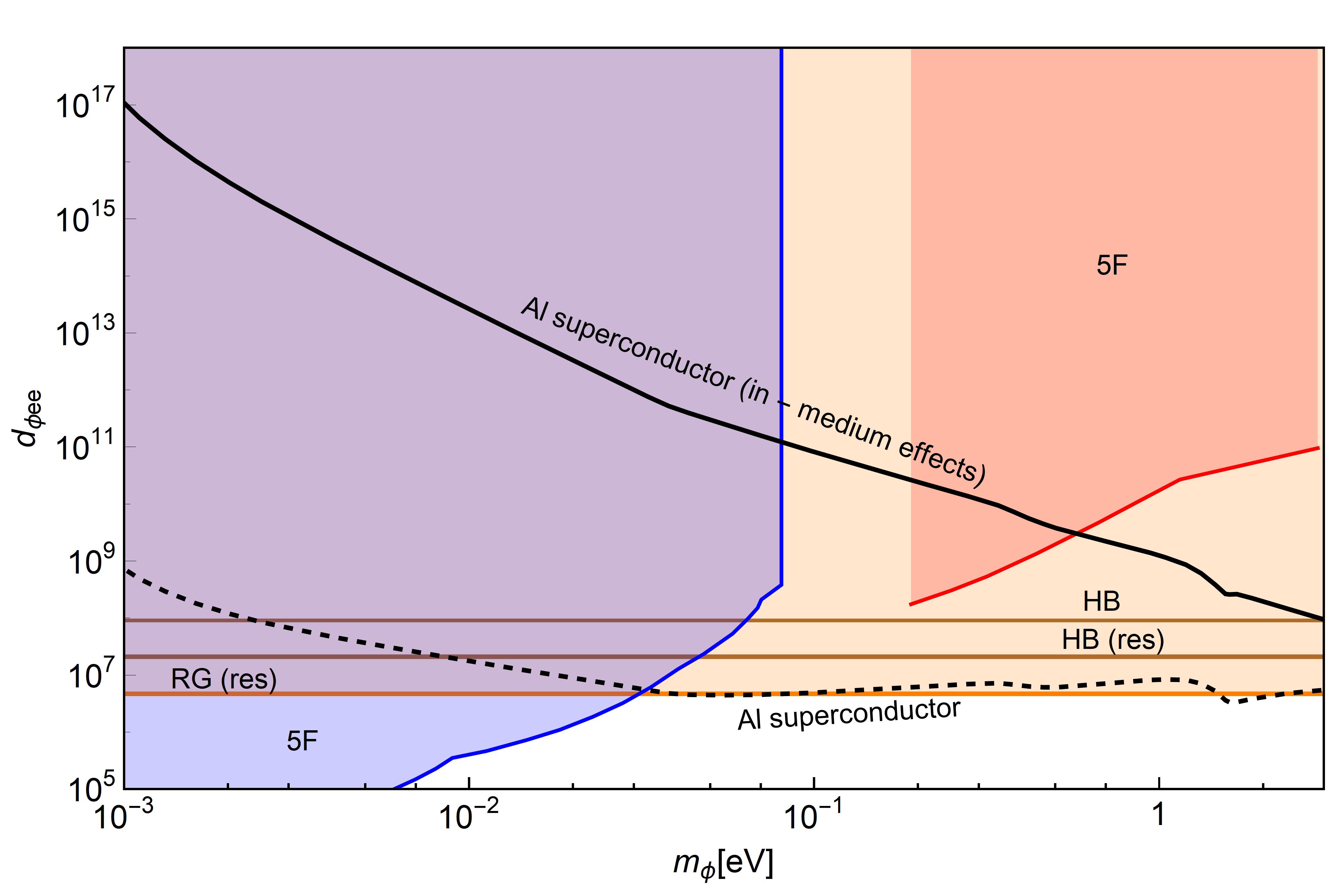}
\caption{ \label{fig:scalarsuper} 
Estimated sensitivity of an aluminum superconductor target for 1-kg-year exposure (left panel) without in-medium effects (dashed black)~\cite{Hochberg:2016ajh} and (right panel) including in-medium effects (solid black), for absorption of scalar DM. We also display constraints from fifth-force searches (shaded blue and red, ``5F'')~\cite{Adelberger:2003zx}, horizontal branch cooling from continuum production (brown line, ``HB'') and resonant production (thick brown line, ``HB (res)'')~\cite{Hardy:2016kme}, red giant resonant production (shaded orange, ``RG (res)'')~\cite{Hardy:2016kme}.}
\end{center}
\end{figure*} 

Here, we follow the discussion of Refs.~\cite{Gelmini:2020kcu, Caputo:2020quz} of
the absorption of axions and dark photons in a detector, respectively.\footnote{The same approach has been previously used to compute the production of particles
in stellar environments, see e.g. Refs.~\cite{Hardy:2016kme, Redondo:2013lna}.} The emission and absorption rates of a boson by a medium are directly related to the self-energy $\Pi$ of the particle in the medium itself as~\cite{Weldon:1983jn, Kapusta:2006pm}
\begin{equation} \label{eq:optic}	
{\rm Im}\,\Pi=-\omega \Gamma\, .
\end{equation}

Here $\omega$ is the particle energy and ${\Gamma=\Gamma_{\rm abs}-\Gamma_{\rm prod}}$ is the rate with which the considered particle momentum distribution approaches thermal equilibrium. Eq.~\eqref{eq:optic} is the optical theorem restated in the framework of thermal field theory. With momentum distribution $f_\phi(\bold{k})$, the scalar number density $n$ is 
 \begin{align}\label{app_numb} 
     n=\int \frac{d^3 \bold{k}}{(2\pi)^3} f_\phi \, .
 \end{align}
For a boson with absorption rate $\Gamma_{\rm abs}$ and production rate $\Gamma_{\rm prod}$, one finds~\cite{Weldon:1983jn}
\begin{equation}
    \frac{\partial f}{\partial t}=-f\Gamma_{\rm abs}+(1+f)\Gamma_{\rm prod}\, .
\end{equation}
 To reduce noise, the detectors relevant for our discussion typically operate at cryogenic temperatures. Hence, we can neglect any population of photons in the medium. Plasmons are only produced by
 the absorption of DM. For scalar masses of $\mathcal{O}(\lesssim 10~\rm{ eV})$ the occupation number of DM bosons is large, $f_\phi\gg 1$, so that the scalar and photon $A$ distributions evolve respectively as
\begin{subequations}
\begin{align}
    \frac{\partial f_\phi}{\partial t}&\simeq-f_\phi\Gamma^{\phi}\\
    \frac{\partial f_A}{\partial t}&=-f_A\Gamma^{\phi}_{\rm prod}+(1+f_A)\Gamma^{\phi}_{\rm abs}\, ,
\end{align}
\end{subequations}
where $\Gamma^{\phi}=\Gamma^{\phi}_{\rm abs}-\Gamma^{\phi}_{\rm prod}$ is the rate by which $\phi$ particles are driven towards equilibrium.

Using $\partial f_A/\partial t=-\partial f_\phi/\partial t$, one has
\begin{equation}\label{ratestep1}
    \Gamma^{\phi}_{\rm abs}\simeq(f_\phi-f_A)\Gamma^{\phi}\simeq -f_\phi\frac{{\rm Im}\Pi_{\phi}}{\omega}\, ,
\end{equation}
where in the last approximation we have assumed that $f_\phi\gg f_A$  and used Eq.~\eqref{eq:optic}.

To lowest order the scalar self-energy is given in the vacuum propagation basis (i.e. the mass basis) by~\cite{Hardy:2016kme}
\begin{align}\label{selfenergy}
	\Pi_{\phi}(K)=\left( \Pi^{\phi\phi} - \frac{(\Pi^{A\phi})^2}{\Pi^{AA} - m_\phi^2}
	\right)~,
\end{align}
where $K=(\omega,\bf{k})$ is the four-momentum of the 
scalar, and $\Pi^{\phi\phi}$, $\Pi^{A\phi}$, and $\Pi^{AA}$ correspond to the forward scattering amplitudes off
electrons for the processes $\phi+e\rightarrow \phi+e$, $A+e\rightarrow \phi+e$, and $A+e\rightarrow A+e$ respectively. 
The second term of Eq.~\eqref{selfenergy} has long been known to be important for dark photon absorption in the detector. However, its effect on scalar absorption has not been previously appreciated. 

At this point, one simply needs expressions for $\Pi^{\phi\phi}$, $\Pi^{A\phi}$ and $\Pi^{AA}$, given in Appendix~A, and 
 substitute Eq.~\eqref{selfenergy} into Eq.~\eqref{ratestep1}.

 The resulting absorption rate of scalars, as shown in Appendix~A, is (see also Ref.~\cite{Hardy:2016kme})
\begin{align} \label{eq:longplasmon}
	\Gamma^{\phi}_{\rm abs} &=f_\phi \frac{g^2}{e^2} k^2\omega^2\frac{ \Gamma_L }{(\omega^2-\omega_p^2)^2+(\omega \Gamma_L)^2}\,  ,
\end{align}
where $\omega_p^2=(\omega^2/K^2)\mathrm{Re}\, \Pi_L $ is the plasma frequency (squared) and $\Gamma_L=-Z_L\mathrm{Im}\, \Pi_L/\omega$ is the damping rate of longitudinal plasmons in the material. The renormalization factor $Z_L=\omega^2/K^2$ is needed for the longitudinal plasmon, see Appendix~A. This 
rate closely resembles the production rate of scalars from a thermal bath of SM particles obtained in Ref.~\cite{Hardy:2016kme}. 
We observe that  $\omega\gg \omega_p$ and $\omega\gg \Gamma_L$ give the relevant limit for in-medium effects to be
negligible. 

The above results imply  that the
in-medium effects can be included by defining an effective coupling of the scalar DM
\begin{align} \label{eq:chieff}
	g_{\rm eff}^2=g^2 \dfrac{m_\phi^4}{(m_\phi^2-\omega_p^2)^2+(m_\phi \Gamma_L)^2} \, ,
\end{align}
where in the DM non-relativistic limit $\omega \simeq m_{\phi}$.
 We thus find that the scalar DM coupling gets modified by in-medium effects just like in the case of dark photon DM, e.g.~\cite{Hochberg:2017wce,Coskuner:2019odd,An:2014twa}.

\section{Scalar absorption in superconductors}

As an example where in-medium effects are important, we consider scalar DM absorption in aluminum superconductors~(see e.g.~\cite{Hochberg:2016ajh}). Following Ref.~\cite{Hochberg:2016ajh},
we take the coupling of Eq.~\eqref{eq:lagrangian} to be $g = d_{\phi ee} \sqrt{4 \pi} (m_e/M_{\rm pl})$.    
The longitudinal component of the polarization tensor $\Pi_L$ is related to the complex index of refraction $\hat{n}$, the complex dielectric function $\hat{\epsilon}$, 
as well as the complex conductivity    
 $\hat{\sigma} = \sigma_1 + i \sigma_2$ of the medium,
 $\Pi_L=K^2(1-\hat{n}^2)=K^2(1-\hat{\epsilon})=- K^2(i\hat{\sigma}/\omega)$. Using $\sigma_1$, 
the scalar absorption rate is given in Ref.~\cite{Hochberg:2016ajh}. Including in-medium effects via coupling redefinition $d_{\phi ee} \rightarrow d_{\phi ee, \rm eff}$ from Eq.~\eqref{eq:chieff} the rate becomes
\begin{align}\label{eq:Rscalar}
	R =&~ \frac{1}{\rho_T}  \frac{\rho_{\rm DM}}{m_\phi}  \frac{3}{\alpha} \left( d_{\phi ee, \rm eff} \frac{m_e}{M_{\rm pl}}  \right)^2 \sigma_1  \nonumber \\
	&~ \times \begin{cases}
 	\dfrac{5}{2}c_s^2 &,\  \  \omega < \omega_D \\    \\
	\dfrac{5}{3} \dfrac{c_s^2 \omega^2}{\omega_D^2} \dfrac{\left(1 - ({3 \omega_D}/{4 \omega}) \right)}{\left(1 - ({5 \omega_D}/{6 \omega}) \right)}   &,  \ \ \omega > \omega_D
	\end{cases}
\end{align}
where $\rho_T$ is the mass density of target material, $\rho_{\rm DM} = 0.3$~GeV/cm$^3$
is the local DM mass density, $m_\phi$ is the DM mass, $m_e$ is the electron mass, $M_{\rm pl}$ is the Planck mass,
$c_s \simeq 2 \times 10^{-5}$ is the sound speed in aluminum, $\omega_D \simeq 0.037$~eV is the maximum frequency for phonons in aluminum and $\sigma_1$ is the real
part of the complex conductivity in aluminum, given in Ref.~\cite{Hochberg:2016ajh},
which corresponds to the physical width $\Gamma_L$ of the longitudinal plasmon (i.e.~$\Gamma_L=  \sigma_1$ for $\omega= m_\phi$).

In Fig.~\ref{fig:scalarsuper} we display the projected sensitivity for scalar DM absorption in an aluminum superconductor for 1-kg-year exposure, along with other existing bounds,
comparing the cases with and without in-medium effects. The 
sensitivity is suppressed by in-medium effects by a factor 
\begin{equation}
    \left(\frac{d_{\phi ee, \rm eff}}{d_{\phi ee}}\right)^2  = \frac{m_{\phi}^4}{(m_{\phi} \sigma_1)^2 + (m_{\phi}^2 - \omega_p^2)^2}~,
\end{equation}
where $\omega_p \simeq 12.2$~eV is the plasma frequency of aluminum. To treat other materials in general, one can substitute $\omega_p^2\rightarrow(\omega^2/K^2) \mathrm{Re}\, \Pi_L $. 

\section{Scalar-mediated dark matter scattering}

Let us now assume that the DM is composed of fermions  $f_{DM}$ that couple to the SM through a scalar portal,
\begin{equation}
	\mathcal{L} \supset  g\phi \bar{e} e+ g_{DM} \phi \bar{f}_{DM} f_{DM}  \, .
\end{equation}
This allows us to determine the scattering rate of DM off electrons through a scalar mediator, which is also affected by in-medium effects. 

To leading order,  the electron-scalar coupling $g$ to be used in scattering computations becomes
\begin{equation}
g_{\mathrm{eff}} = g\dfrac{  q^2}{ (q^2 - \Pi_L) }~.
\end{equation}
where $q$ is the magnitude of the momentum transfer space component. One can diagonalize the quadratic terms in the Lagrangian, i.e. the matrix in Eq.~(2.1) of Ref.~\cite{Hardy:2016kme}, where $K$ is now the momentum transfer four-vector. Inverting the eigenvalue corresponding to the eigenvector with the largest scalar component, one finds that neglecting terms of $\mathcal{O} (g^2)$, the scalar propagator is the same as in vacuum.  Thus, the DM-electron scattering amplitude for non-relativistic DM and electrons (i.e., where the Pauli matrices are replaced by $\gamma^0=1$, $\gamma^i=0$) is
\begin{equation}
 g_{DM}~ g~ \bar{u}_{DM}~ {u}_{DM} \dfrac{1}{(K^2-m_\phi^2)} \dfrac{q^2}{ (q^2 - \Pi_L) }   \bar{u}_e  {u}_e~,  
\end{equation}
where $u$ is the usual Dirac spinor.

This amplitude coincides with the dark photon mediated scattering amplitude, given in e.g. Eqs.~(5.10) and (5.15) of Ref.~\cite{Hochberg:2015fth}, if we replace $g\rightarrow \epsilon e$. Therefore, for example, the experimental sensitivities in Fig.~8 of Ref.~\cite{Hochberg:2017wce} (for a light scalar mediator) should be replaced by those in Fig.~4 of the same reference (for a dark photon mediator), where also $g= \epsilon e$.

\section{Mitigating in-medium effects}

The in-medium effects described above 
potentially affect the absorption of light scalar DM in a variety of materials. Besides superconductors, as in the example above, semiconductors
(e.g.~\cite{Hochberg:2016ntt}) or polar materials \cite{Knapen:2017ekk} could also have sensitivities different
than expected.  In multi-tonne direct detection experiments 
based on xenon  or argon,  searches of scalar halo DM with mass of $\mathcal{O}$(eV) focusing on an ionization signal could also be affected, in analogy to dark photon searches~(e.g.~\cite{An:2014twa}). 

In-medium effects are negligible if one is far from resonance, i.e. $\omega\gg \omega_p$ and $\omega\gg \Gamma_L$.  While halo DM particles have very low characteristic speeds of $\sim 10^{-3}$, light scalars could be produced with much larger kinetic energies from particular sources, e.g. in the Sun~\cite{Budnik:2019olh}, in cosmic-ray interactions~\cite{Bringmann:2018cvk,Plestid:2020kdm,Ema:2018bih} or in the decay of other DM particles~\cite{Agashe:2014yua}. 
Searches of scalars in multi-tonne xenon and argon experiments
based on  electron recoil energies of $\mathcal{O}$(keV), as e.g. in Ref.~\cite{Aprile:2020tmw}, are not significantly affected by in-medium effects, since $\omega \gg \sigma_1,\sigma_2$. 

We note that in-medium effects can not only suppress a signal but also enhance it,
taking advantage  of the associated resonance. 
This is done by matching 
 the resonance frequency with the particle mass, i.e. $\omega_p= m_\phi$, as accomplished with tunable plasma haloscopes~\cite{Lawson:2019brd,Gelmini:2020kcu}. 
 
In-medium effects can also be mitigated for scalar mediated DM scattering by a judicious  choice of target materials.

\section{Conclusions}
 
We have studied here important previously ignored in-medium effects for the absorption of scalars in
detector materials. These effects suppress the signal when the scalar energy is not much larger than the plasma frequency $\omega_p$ of the medium. The same effects can, on the contrary, enhance the signal through a resonant absorption in materials or metamaterials, in which $\omega_p$ can be tuned to be equal to the scalar energy.  

We have computed the in-medium effects for scalar detection using thermal field theory techniques, and have shown that the reach of several proposed direct detection experiments searching for light halo DM scalars is significantly weakened. 

We also discussed the effect of the medium on scalar mediated DM scattering, finding a suppression identical to that of a dark photon mediated scattering for non-relativistic DM and electrons.

These developments identify setups, such as those based on Dirac materials and tunable plasma haloscopes, as particularly promising for scalars in direct detection.

 We have also shown that for scalars with energy much larger than $\mathcal{O}(10~\rm{eV})$, e.g.  such as those produced in the Sun, cosmic rays or decays of other DM particles,  in-medium effects are negligible. 
Hence, experiments searching for a keV-level ionization signal, such as those based on xenon or argon, also constitute a favorable target.

\section*{Acknowledgments}
We thank G.~Raffelt for discussions.
The work of GG, VT and EV was supported by the U.S. Department of Energy (DOE) Grant No. DE-SC0009937.

\appendix

\section*{Appendix A: Particle self-energies in plasma}
\label{tftapp}
Here, we review the thermal field theory machinery needed to compute the self-energies of particles in a plasma. We will primarily follow the treatment of Ref.~\cite{Hardy:2016kme}, where the mixing of scalars and longitudinal plasmons has been discussed in the context of astrophysical environments. This approach is well known and has been previously applied to the production of other particles in stars, like dark photons~(e.g. \cite{An:2013yfc,Redondo:2013lna}) and neutrinos~(e.g. \cite{Braaten:1993jw,Haft:1993jt}).
In-medium interactions can be accounted for by including a linear response (see, for instance, Refs.~\cite{Raffelt:1996wa,Haft:1993jt}), 
\begin{align}
    J^\mu_{\rm ind}=-\Pi^{\mu\nu}{A}_\nu \, ,
\end{align}
where $\Pi^{\mu\nu}$ is a polarization tensor. Hence, the total current coupling to photons is $ J_{\rm EM}^{\mu} + J^\mu_{\rm ind}$. 

In the Lorenz gauge, the polarization tensor in an isotropic medium is described by two polarization functions $\Pi_T$ and $\Pi_L$, for transverse and longitudinal excitations (e.g.~\cite{An:2014twa}),
\begin{align}
    \Pi^{\mu\nu}\equiv ie^2\langle J^\mu_{\rm EM}J^\nu_{\rm EM}\rangle&=-\Pi_T (e_+^\mu e_+^{*\nu}+e_-^\mu e_-^{*\nu})-
    \Pi_L e_L^\mu e_L^{*\nu}
    \notag
    \\
    &=\sum_{i=\pm, L} \Pi_i P_i^{\mu\nu}~,
\end{align}
where $e_L$ and $e_{\pm}$ are the longitudinal and the transverse polarization vectors and $P_i^{\mu\nu}$ are projectors. Assuming $\bold{k}$ is parallel to the $\hat{z}$ axis, the polarization vectors are
\begin{align}
e_L\equiv\frac{(k^2,\omega\bold{k})}{k\sqrt{K^2}} \quad{\rm and}\quad 
e_{\pm}\equiv\frac{1}{\sqrt{2}}(0,\bold{e}_x\pm i \bold{e}_y) \, ~,
\end{align}
where $\bold{e}_x, \bold{e}_y$ are orthogonal  unit vectors perpendicular to the unit vector $\bold{k}/k$.

To lowest order, the real part of the polarization tensor is obtained from the forward scattering on charged particles. Moreover, because the scattering amplitude involves for a non-relativistic plasma the inverse mass of the targets, we can limit our attention to electrons. Thus, the real part of the polarization tensor is~\cite{Braaten:1993jw, Raffelt:1996wa}
\begin{align}
	\rm{Re~}&\Pi^{\mu\nu} (K) = 4e^2 \int \frac{d^3 p}{(2 \pi)^3} \frac{1}{2 E_p}
	(f_e (E_p) + f_{\bar{e}}(E_p)) \nonumber \\
	&\times\frac{ K^2 P^\mu
	P^\nu +(P \cdot K)^2 g^{\mu\nu}-P \cdot K (P^\mu K^\nu + K^\mu P^\nu) }{(P \cdot K)^2 - (K^2)^2 / 4} ~,
	\label{eq:piAA}
\end{align}
which is correct to order $\alpha$. Here, $f_e$ and $f_{\bar{e}}$ are the distribution functions for electrons and positrons. In the non-relativistic limit, one considers only electrons.
As discussed in Refs.~\cite{Braaten:1993jw, Raffelt:1996wa},
the $K^4$ term in the denominator can be neglected.

Ignoring the $K^4$ term and multiplying  by the projectors, in the lowest order of electron velocities one obtains~\cite{Redondo:2013lna,Raffelt:1996wa}
 \begin{subequations}
\begin{alignat}{2}
	{\rm Re}\, \Pi_{T}&=\omega_p^2\, ,\\
	{\rm Re}\,\Pi_{L}&=\frac{K^2}{\omega^2}\omega_p^2\, .
\end{alignat}
\end{subequations}
The dispersion relation for the transverse plasmon is $\omega^2~-~k^2=~\omega_p^2$. This allows for an interpretation of transverse excitations as particles with mass $\omega_p$. In a plasma we have $\omega_p=e^2 n_e/m_e$, while in a metal one has to use the effective mass of electrons $m_e^*$. Longitudinal plasmons instead have a dispersion relation given by $\omega^2=\omega_p^2$. 

The imaginary part of the photon polarization tensor can be potentially computed via Kramers-Kronig relations, or experimentally measured. The transverse plasmon absorption is given for a non-relativistic plasma by $
    \Gamma_T=-{\rm Im}\,\Pi_{T}/\omega \,$.
When treating longitudinal plasmons one defines the vertex renormalization constant $Z_L=\omega^2/K^2$, which must be included for external longitudinal plasmon in a diagram. In this way, the longitudinal rate is $
    \Gamma_L=-Z_L{\rm Im}\,\Pi_{L}/\omega \,$.
 \\
 
We can now turn our focus on the scalar self-energy computation. The one-loop electron self-energies, assuming the Lagrangian of Eq.~\eqref{eq:lagrangian}, are
\begin{align}
\mathrm{Re}\, \Pi^{\phi A,\mu} (K) =&~ 4g_\phi e \int \frac{d^3 p}{(2 \pi)^3}  \frac{1}{2 E_p}
	(f_e (E_p) + f_{\bar{e}}(E_p)) \nonumber \\
	&\times  m_e \frac{K^2 P^\mu-(P \cdot K) K^\mu }{(P \cdot K)^2 - (K^2)^2 / 4} ~,
\end{align}
for the scalar-photon mixing and 
\begin{align}
	\mathrm{Re}\, \Pi^{\phi\phi} (K) =&~ 4g_\phi^2 \int \frac{d^3 p}{(2 \pi)^3}  \frac{1}{2 E_p}
	(f_e (E_p) + f_{\bar{e}}(E_p)) \nonumber \\
	&~\times  \frac{(P \cdot K)^2 - m_e^2 K^2}{(P \cdot K)^2 - (K^2)^2/4} ~,
\end{align}
for the scalar-scalar mixing. In a non-relativistic plasma, these give
\begin{align}\label{piphiLL}
    \mathrm{Re}~\Pi^{\phi L}&=\mathrm{Re}~\Pi^{\phi A,\mu}e^L_\mu\notag \\ &\simeq g e \frac{k \sqrt{K^2}}{\omega^2} \frac{n_e}{m_e} \, = \frac{g}{e}    \frac{k}{\sqrt{K^2}}\, \mathrm{Re}\,\Pi^L
\end{align}
and
\begin{equation}\label{piphiphiphi}
    \mathrm{Re}~\Pi^{\phi \phi}\simeq \frac{g^2}{e^2}    \frac{k^2}{K^2}\mathrm{Re}\,\Pi^L \ .
\end{equation}

To lowest order the total scalar self-energy is given by~\cite{Hardy:2016kme}
\begin{align}\label{appselfenergy}
	\Pi_{\phi}(K)=\left( \Pi^{\phi\phi} - \frac{(\Pi^{A\phi})^2}{\Pi^{AA} - m_\phi^2}
	\right)~,
\end{align}
where $K=(\omega,\bf{k})$ is the four-momentum of the external scalar, and $\Pi^{\phi\phi}$, $\Pi^{A\phi}$, and $\Pi^{AA}$ correspond to the forward scattering amplitudes over electrons for the processes $\phi+e\rightarrow \phi+e$, $A+e\rightarrow \phi+e$, and $A+e\rightarrow A+e$ respectively. One then finds
\begin{align}
	\mathrm{Im}\,\Pi^\phi~=&~	\mathrm{Im}\,\Pi^{\phi\phi} \notag\\
	&+ \frac{\mathrm{Im}\,\Pi^{AA} ((\mathrm{Re}\,\Pi^{A\phi})^2 - (\mathrm{Im}\,\Pi^{A\phi})^2)}{(\mathrm{Im}\,\Pi^{AA})^2 + (\mathrm{Re}\,\Pi^{AA} - m_\phi^2)^2}\notag 
	\\
	&-\frac{
	 2 (\mathrm{Re} \, \Pi^{AA} - m_\phi^2)\,  \mathrm{Re} \,\Pi^{A\phi} \mathrm{Im} \,\Pi^{A\phi}}{(\mathrm{Im} \,\Pi^{AA})^2 + (\mathrm{Re} \, \Pi^{AA} - m_\phi^2)^2} \, .
\end{align}

Enforcing the on-shell condition $K^2=m_\phi^2$, the resulting absorption rate of scalars at leading order is (see also Ref.~\cite{Hardy:2016kme})
\begin{align} \label{eq:longplasmon}
	\Gamma^{\phi}_{\rm abs} &=f_\phi \frac{g^2}{e^2} k^2\omega^2\frac{ \Gamma_L }{(\omega^2-\omega_p^2)^2+(\omega \Gamma_L)^2}\,  .
\end{align}
This equation can be obtained using Eqs.~\eqref{piphiLL} and~\eqref{piphiphiphi}, together with the equivalent expressions for the imaginary parts derived using unitarity.
The absorption rate of scalars per unit target mass in a detector is thus
\begin{align}\label{rateedo}
    R&=\frac{1}{\rho_T}\int \frac{d^3\bold{k}}{(2\pi^3)}\Gamma^\phi_\mathrm{abs} \\
    &=\frac{1}{\rho_T}\int \frac{d^3\bold{k}}{(2\pi^3)}f_\phi \frac{g^2}{e^2} k^2 m_\phi^2\frac{ \Gamma_L }{(m_\phi^2-\omega_p^2)^2+(m_\phi \Gamma_L)^2} \, , \notag
\end{align}
where $\rho_T$ is the detector mass density. As this equation shows, the DM absorption rate can be directly related to the photon absorption rate specified via $\Gamma_L$, which is experimentally measured.

\section*{Appendix B: Absorption in superconductors}
\label{app:supercond}

We summarize below the treatment of the scalar absorption rate in superconductors following  Ref.~\cite{Hochberg:2016ajh}. The problem can be described in terms of macroscopic quantities and the results matched into particle physics model parameters via Feynman diagram calculations. The general DM-electron absorption rate is given by
\begin{align}
    R&=\frac{1}{\rho_T}\int \frac{d^3\bold{k}}{(2\pi^3)}\Gamma^\phi_\mathrm{abs}\notag\\
    &=\frac{1}{\rho_T}\frac{\rho_{\rm DM}}{m_{\rm DM}}\langle n_e \sigma_{\rm abs}v_{\rm rel}\rangle \, ,
\end{align}
where $\sigma_{\rm abs}$ is the absorption cross-section, $n_e$ is the electron number density, and the average is over DM velocity $v_{\rm rel}$.

 In the limit of small momentum $k$ compared to the energy $\omega$, $k\ll \omega$, the transverse and longitudinal parts of the polarization tensor characterizing the medium are equal and are related to the complex conductivity as
\begin{equation}
        \Pi(\omega) \simeq \Pi_T(\omega) \simeq \Pi_L(\omega) \simeq - i \hat{\sigma} \omega~.
\end{equation}

From the optical theorem, the photon absorption rate can be related to the polarization tensor via
\begin{equation}
   \langle n_e \sigma_{\rm abs}v_{\rm rel}\rangle_{\gamma} = \Gamma_L=- \frac{\operatorname{Im}\Pi(\omega)}{\omega}~.
\end{equation}  
The absorption rate is characterized by $\sigma_1$, with $\sigma_1 = \langle n_e \sigma_{\rm abs}v_{\rm rel}\rangle_{\gamma}$.

Using the standard Drude model for metals to describe the conductivity, one obtains
\begin{align}
    \sigma_1(\omega) \simeq \frac{\omega_p^2}{\omega^2 \tau} \qquad,\qquad
    \sigma_2(\omega) \simeq \frac{\omega_p^2}{\omega} \, ,
\end{align}
where $\tau$ is the electron scattering time in a medium and in aluminum $\omega_p \simeq 12.2$~eV. In the above, we assumed that $\omega \tau \gg 1$ and also that $\tau$ does not depend on temperature, envisioning a cryogenic system.

Here, the interaction time $\tau$ is set by athermal phonons interacting with electrons. Using the Debye model for the phonon dispersion, one obtains
\begin{align}
	\dfrac{1}{\tau} =  \begin{cases}
 	\dfrac{4}{5} \pi \lambda_{\rm tr} \omega_D ( 1 - \dfrac{5}{6}\dfrac{\omega_D}{\omega}) &,\ \  \omega \geq \omega_D \\  \\
	\dfrac{2}{15} \pi \lambda_{\rm tr} \dfrac{\omega^5}{\omega_D^4}   &,  \ \ \omega < \omega_D
	\end{cases}
\end{align}
where for aluminum $\omega_D \simeq 0.037$~eV and the measured high temperature resistivity is $\lambda_{\rm tr} = 0.39$~\cite{PhysRevB.36.2920}.

While the Drude model is in general only valid for metals in a non-superconducting phase, the difference appears close to the superconducting gap $2 \Delta$. This difference can be accommodated as in Ref.~\cite{PhysRevB.3.305} assuming the electron scattering time $\tau_{\rm super}$ in the superconductor is such that $1/\tau_{\rm super} = f(\Delta)/\tau$, where $f(\Delta)$ is an integral function of the gap. This only affects the results when $\omega \lesssim 10^{-2}~\mathrm{eV}$~\cite{Hochberg:2016ajh}.

Matching the matrix elements $|\mathcal{M}|^2 \propto |\mathcal{M}_{\gamma}|^2$ for scalar and photon interactions and using the macroscopic quantities discussed above, one obtains the absorption rate for scalars as described in Eq.~\eqref{eq:Rscalar}. For our approximate results in this manuscript, we employ the $\sigma_1$ obtained in Ref.~\cite{Hochberg:2016ajh} for normal metals (see their Fig.~2).

\bibliographystyle{bibi}
\bibliography{bibliography}
 
\end{document}